

\def\singlespace{\normalbaselines}
\def\oneandahalfspace{\baselineskip=1.15\normalbaselineskip plus 1pt
\lineskip=2pt\lineskiplimit=1pt}

\def\np{\vfill\eject}
\def\nl{\hfil\break}

\def\nofirstpagenoten{\nopagenumbers\footline={\ifnum\pageno>1\tenrm
\hss\folio\hss\fi}}
\def\nofirstpagenotwelve{\nopagenumbers\footline={\ifnum\pageno>1\twelverm
\hss\folio\hss\fi}}
\def\leaderfill{\leaders\hbox to 1em{\hss.\hss}\hfill}
\def\ft#1#2{{\textstyle{{#1}\over{#2}}}}
\def\frac#1/#2{\leavevmode\kern.1em
\raise.5ex\hbox{\the\scriptfont0 #1}\kern-.1em/\kern-.15em
\lower.25ex\hbox{\the\scriptfont0 #2}}
\def\sfrac#1/#2{\leavevmode\kern.1em
\raise.5ex\hbox{\the\scriptscriptfont0 #1}\kern-.1em/\kern-.15em
\lower.25ex\hbox{\the\scriptscriptfont0 #2}}


\parindent=20pt
\def\narrow{\advance\leftskip by 40pt \advance\rightskip by 40pt}

\def\AB{\bigskip
        \centerline{\bf ABSTRACT}\medskip\narrow}
\def\nonarrower{\advance\leftskip by -40pt\advance\rightskip by -40pt}
\def\AE{\bigskip\nonarrower}

\def\boxit#1{\vbox{\hrule\hbox{\vrule\kern3pt
        \vbox{\kern3pt#1\kern3pt}\kern3pt\vrule}\hrule}}

\def\gtorder{\mathrel{\raise.3ex\hbox{$>$}\mkern-14mu
             \lower0.6ex\hbox{$\sim$}}}
\def\ltorder{\mathrel{\raise.3ex\hbox{$<$}|mkern-14mu
             \lower0.6ex\hbox{\sim$}}}
\def\dalemb#1#2{{\vbox{\hrule height .#2pt
        \hbox{\vrule width.#2pt height#1pt \kern#1pt
                \vrule width.#2pt}
        \hrule height.#2pt}}}

\font\fourteentt=cmtt10 scaled \magstep2
\font\fourteenbf=cmbx12 scaled \magstep1
\font\fourteenrm=cmr12 scaled \magstep1
\font\fourteeni=cmmi12 scaled \magstep1
\font\fourteenss=cmss12 scaled \magstep1
\font\fourteensy=cmsy10 scaled \magstep2
\font\fourteensl=cmsl12 scaled \magstep1
\font\fourteenex=cmex10 scaled \magstep2
\font\fourteenit=cmti12 scaled \magstep1
\font\twelvett=cmtt10 scaled \magstep1 \font\twelvebf=cmbx12
\font\twelverm=cmr12 \font\twelvei=cmmi12
\font\twelvess=cmss12 \font\twelvesy=cmsy10 scaled \magstep1
\font\twelvesl=cmsl12 \font\twelveex=cmex10 scaled \magstep1
\font\twelveit=cmti12
\font\tenss=cmss10
 
 \font\ninebf=cmbx7 scaled \magstep1
\font\ninerm=cmr7 scaled \magstep1 \font\ninei=cmmi7 scaled \magstep1
\font\ninesy=cmsy7 scaled \magstep1 
\font\eightrm=cmr7 scaled 1140 
 
\font\sevenbf=cmbx7 \font\sevenrm=cmr7 \font\seveni=cmmi7
\font\sevensy=cmsy7 

\catcode`@=11
\newskip\ttglue
\newfam\ssfam

\def\fourteenpoint{\def\rm{\fam0\fourteenrm}
\textfont0=\fourteenrm \scriptfont0=\tenrm \scriptscriptfont0=\sevenrm
\textfont1=\fourteeni \scriptfont1=\teni \scriptscriptfont1=\seveni
\textfont2=\fourteensy \scriptfont2=\tensy \scriptscriptfont2=\sevensy
\textfont3=\fourteenex \scriptfont3=\fourteenex \scriptscriptfont3=\fourteenex
\def\it{\fam\itfam\fourteenit} \textfont\itfam=\fourteenit
\def\sl{\fam\slfam\fourteensl} \textfont\slfam=\fourteensl
\def\bf{\fam\bffam\fourteenbf} \textfont\bffam=\fourteenbf
\scriptfont\bffam=\tenbf \scriptscriptfont\bffam=\sevenbf
\def\tt{\fam\ttfam\fourteentt} \textfont\ttfam=\fourteentt
\def\ss{\fam\ssfam\fourteenss} \textfont\ssfam=\fourteenss
\tt \ttglue=.5em plus .25em minus .15em
\normalbaselineskip=16pt
\abovedisplayskip=16pt plus 4pt minus 12pt
\belowdisplayskip=16pt plus 4pt minus 12pt
\abovedisplayshortskip=0pt plus 4pt
\belowdisplayshortskip=9pt plus 4pt minus 6pt
\parskip=5pt plus 1.5pt
\setbox\strutbox=\hbox{\vrule height12pt depth5pt width0pt}
\let\sc=\tenrm
\let\big=\fourteenbig \normalbaselines\rm}
\def\fourteenbig#1{{\hbox{$\left#1\vbox to12pt{}\right.\n@space$}}}

\def\twelvepoint{\def\rm{\fam0\twelverm}
\textfont0=\twelverm \scriptfont0=\ninerm \scriptscriptfont0=\sevenrm
\textfont1=\twelvei \scriptfont1=\ninei \scriptscriptfont1=\seveni
\textfont2=\twelvesy \scriptfont2=\ninesy \scriptscriptfont2=\sevensy
\textfont3=\twelveex \scriptfont3=\twelveex \scriptscriptfont3=\twelveex
\def\it{\fam\itfam\twelveit} \textfont\itfam=\twelveit
\def\sl{\fam\slfam\twelvesl} \textfont\slfam=\twelvesl
\def\bf{\fam\bffam\twelvebf} \textfont\bffam=\twelvebf
\scriptfont\bffam=\ninebf \scriptscriptfont\bffam=\sevenbf
\def\tt{\fam\ttfam\twelvett} \textfont\ttfam=\twelvett
\def\ss{\fam\ssfam\twelvess} \textfont\ssfam=\twelvess
\tt \ttglue=.5em plus .25em minus .15em
\normalbaselineskip=14pt
\abovedisplayskip=14pt plus 3pt minus 10pt
\belowdisplayskip=14pt plus 3pt minus 10pt
\abovedisplayshortskip=0pt plus 3pt
\belowdisplayshortskip=8pt plus 3pt minus 5pt
\parskip=3pt plus 1.5pt
\setbox\strutbox=\hbox{\vrule height10pt depth4pt width0pt}
\let\sc=\ninerm
\let\big=\twelvebig \normalbaselines\rm}
\def\twelvebig#1{{\hbox{$\left#1\vbox to10pt{}\right.\n@space$}}}

\def\tenpoint{\def\rm{\fam0\tenrm}
\textfont0=\tenrm \scriptfont0=\sevenrm \scriptscriptfont0=\fiverm
\textfont1=\teni \scriptfont1=\seveni \scriptscriptfont1=\fivei
\textfont2=\tensy \scriptfont2=\sevensy \scriptscriptfont2=\fivesy
\textfont3=\tenex \scriptfont3=\tenex \scriptscriptfont3=\tenex
\def\it{\fam\itfam\tenit} \textfont\itfam=\tenit
\def\sl{\fam\slfam\tensl} \textfont\slfam=\tensl
\def\bf{\fam\bffam\tenbf} \textfont\bffam=\tenbf
\scriptfont\bffam=\sevenbf \scriptscriptfont\bffam=\fivebf
\def\tt{\fam\ttfam\tentt} \textfont\ttfam=\tentt
\def\ss{\fam\ssfam\tenss} \textfont\ssfam=\tenss
\tt \ttglue=.5em plus .25em minus .15em
\normalbaselineskip=12pt
\abovedisplayskip=12pt plus 3pt minus 9pt
\belowdisplayskip=12pt plus 3pt minus 9pt
\abovedisplayshortskip=0pt plus 3pt
\belowdisplayshortskip=7pt plus 3pt minus 4pt
\parskip=0.0pt plus 1.0pt
\setbox\strutbox=\hbox{\vrule height8.5pt depth3.5pt width0pt}
\let\sc=\eightrm
\let\big=\tenbig \normalbaselines\rm}
\def\tenbig#1{{\hbox{$\left#1\vbox to8.5pt{}\right.\n@space$}}}
\let\rawfootnote=\footnote \def\footnote#1#2{{\rm\parskip=0pt\rawfootnote{#1}
{#2\hfill\vrule height 0pt depth 6pt width 0pt}}}

\def\tenfoot{\tenpoint\hskip-\parindent\hskip-.1cm}

\overfullrule=0pt
\tenpoint
\def\sbullet{\raise.2em\hbox{$\scriptscriptstyle\bullet$}}
\nofirstpagenotwelve
\hsize=16.5 truecm
\baselineskip 15pt

\def\bfbeta{\beta\mkern-10.4mu\beta\mkern-11.2mu\beta\mkern-10.5mu\beta}
\def\bfpsi{\psi\mkern-11.7mu\psi\mkern-11.8mu\psi}
\def\bfphi{\phi\mkern-10.8mu\phi\mkern-10.7mu\phi\mkern-10mu\phi}
\def\bfrho{\rho\mkern-9.9mu\rho\mkern-10.4mu\rho}
\def\bfgamma{\gamma\mkern-10.3mu\gamma\mkern-11mu\gamma
\mkern-10.5mu\gamma}

\def\ft#1#2{{\textstyle{{#1}\over{#2}}}}

\def\a{\alpha_0}

\def\del{\partial}

\def\ket#1{\big| #1\big\rangle}

\oneandahalfspace
\rightline{CTP TAMU--29/92}
\rightline{May 1992}

\vskip 2truecm
\centerline{\bf Aspects of $N=2$ Super-$W_n$ Strings}
\vskip 1.5truecm
\centerline{H. Lu,$^\star$ C.N. Pope,\footnote{$^
\star$}{\tenfoot \sl Supported in part by the
U.S. Department of Energy, under
grant DE-FG05-91ER40633.} X.J. Wang and K.W.
Xu}
\vskip 1.5truecm
\centerline{\it Center
for Theoretical Physics,
Texas A\&M University,}
\centerline{\it College Station, TX 77843--4242, USA.}

\vskip 1.5truecm
\AB\singlespace

      We construct $N=2$ super-$W_{n+1}$ strings and obtain the complete
physical spectrum, for arbitrary $n \ge 2$.  We also derive more general
realisations of the super-$W_{n+1}$ algebras in terms of $k$ commuting $N=2$
super energy-momentum tensors and $n-k$ pairs of complex superfields, with
$0\le k \le [\ft{n+1}{2}]$.

\AE\oneandahalfspace

\vskip 2.0truecm
\centerline{\tenfoot  Available from hep-th/9205054 }

\np
\noindent
{\bf 1. Introduction}
\bigskip

     It is well known that two-dimensional gravity is the gauge theory of
the Virasoro algebra. There are various extensions of the Virasoro algebra,
including the super Virasoro algebras, the $W$ algebras, and the super-$W$
algebras. The gauge theories of these algebras give rise to extensions of
two dimensional gravity. Such a theory is anomalous at the quantum level
unless the central charge of the algebra takes a specific value which
cancels the contribution from the ghosts. Since the ghosts for a bosonic
(fermionic) current of spin $s$ contribute $\mp 2(6 s^2-6s+1)$ to the
central charge, it follows that for an algebra with currents of spin $s$,
the central charge for the matter realisation of the corresponding
$W$-gravity theory is
$$
c=2\sum_{\{s\}_B} \Big(6s^2-6s+1\Big)-2\sum_{\{s\}_F} \Big(6s^2-6s+1\Big)
\ .\eqno(1.1)
$$
Here $\{s\}_B$ and $\{s\}_F$ denote the set of spins of the bosonic and
fermionic currents in the algebra.

     If the matter realisation of the $W$ algebra includes scalar fields,
these fields may be interpreted as the coordinates of spacetime in a
corresponding $W$-string theory. The condition of anomaly freedom (1.1) can
be derived from the nilpotency requirement for the BRST operator. The
intercepts for the bosonic currents are determined by the requirement that
the product of a physical state with the ghost vacuum should be BRST
invariant. Owing to the nonlinearity of the $W$ algebra, this is a
complicated calculation in general. However, since the Virasoro subalgebra
is linear, the intercept for $L_0$ can in fact be determined
straightforwardly. The structure of the ghost vacuum was obtained in general
in [1]; the resulting spin-2 intercept is
$$
\eqalign{
L_0&=\ft12\sum_{\{s\}_B}s(s-1)
-\ft12\sum_{\{s\}_F} \big(s-\ft12\big)^2\cr
&=\ft1{24}\big(c-2N_B-N_F\big)\cr}\ ,\eqno(1.2)
$$
where $N_B$ and $N_F$ are the numbers of bosonic and fermionic currents in
the $W$ algebra. This result gives the conformal dimension of physical
states in the $W$-string theory.

      The bosonic $W_{n+1}$ algebra has one current of each spin $s$ in
interval $2\le s\le {n+1}$, and so
$$
\eqalignno{
c&=2n(2n^2+6n+5)\ ,&(1.3a)\cr
L_0&=\ft16 n(n+1)(n+2)\ .&(1.3b)\cr}
$$
The $W_{n+1}$ algebra here is based on the Miura transformation for the
$A_n$ algebra. In fact any semi-simple Lie algebra $G$ gives rise to an
associated $W$ algebra, denoted by $W\!G$ [2,3]. The $W\!D_n$ algebra has
currents with spins $2,\ 4,\ 6,\ \ldots,\ 2n-2$ and $n$. This implies
$$
\eqalignno{
c&=2n(8n^2-12n+5)\ ,&(1.4a)\cr
L_0&=\ft13 n(n-1)(2n-1)\ .&(1.4b)\cr}
$$
The $W\!B_n$ algebra has bosonic currents with spins $2,\ 4,\ 6,\ \ldots,\
2n$ and a fermionic current with spin $n+\ft12$, which implies
$$
\eqalignno{
c&=(2n+1)(8n^2-4n+1)\ ,&(1.5a)\cr
L_0&=\ft16 n(4n^2-1)\ .&(1.5b)\cr}
$$

    The $N=2$ super-$W_{n+1}$ algebra contains currents which can be grouped
into $N=2$ supermultiplets with the spin content [4]
 $$
\left\{\matrix{&\ft32&\cr
               1&&2\cr
               &\ft32&\cr}\right\}
\left\{\matrix{&\ft52&\cr
               2&&3\cr
               &\ft52&\cr}\right\}
\left\{\matrix{&\ft72&\cr
               3&&4\cr
               &\ft72&\cr}\right\}
\cdots
\left\{\matrix{&(n+\ft12)&\cr
               n&&(n+1)\cr
               &(n+\ft12)&\cr}\right\}\ .\eqno(1.6)
$$
Each lozenge contributes 6 to central charge $c$ and 0 to the intercept
$L_0$. So for the $N=2$ super-$W_{n+1}$ algebra we have
$$\eqalignno{
c&= 6n\ ,&(1.7a)\cr
L_0&=0 \ .&(1.7b)\cr}
$$

    The $N=1$ super-$W_{n+1}$ algebra is a subalgebra of the $N=2$
super-$W_{n+1}$ algebra, whose spin content comprises the even-integer
bosonic spin and one of the fermionic spins in each lozenge [4], namely
$$
\ft32,\ 2,\ 2,\ \ft52,\ \ft72,\ 4,\ \cdots,\ (\ft{n+1}2+[\ft{n+1}2])
\ ,\eqno(1.8)
$$
where $[\ft{n+1}2]$ stands for the integer part of $\ft{n+1}2$. For example,
when $n=1$ we have spins $\ft32,\ 2$; when $n=2$ we have spins $\ft32,\ 2,\
2,\ \ft52$; and when $n=3$ we have spins $\ft32,\ 2,\ 2,\ \ft52,\ \ft72,\
4$. The central charge and the $L_0$ intercept for $N=1$ super-$W_{n+1}$
algebra are
$$
\eqalignno{
c&=12(-1)^{n+1}\ \Big[{{n+1}\over2}\Big]+3n\ ,&(1.9a)\cr
L_0&=\ft12 (-1)^{n+1}\ \Big[{{n+1}\over2}\Big]\ .&(1.9b)\cr}
$$
Note that when $n$ is odd, both $c$ and $L_0$ are positive, whilst they are
both negative when $n$ is even.

      So far, $W$ string theories based on the $W\!A_n$ [5,6,7], $W\!B_n$
and $W\!D_n$ [8,9] algebras have been studied in detail. It is interesting
to study the supersymmetric extensions, and in this paper we shall look at
the $N=2$ super-$W_{n+1}$ algebras and construct the corresponding string
theories. We shall obtain the complete spectrum of the $N=2$ super-$W_{n+1}$
string, and discuss the relation with the $N=2$ super-Virasoro minimal
models. In section 2, we take the known Miura transformation for the
super-$W_{n+1}$ algebra [10,11] and use it to prove that the currents of
super-$W_{n+1}$ may be written in terms of those of super-$W_n$ and an extra
pair of complex superfield. Applying this recursively, we obtain a
realisation in terms of an arbitrary super energy-momentum tensor together
with $(n-1)$ additional pairs of complex superfields. In section 3 we use
this realisation to build a super-$W_{n+1}$ string theory, and obtain its
physical spectrum. In section 4 we discuss the issue of unitarity. Although
some physical states have negative norm, we argue that a truncation to a
subspace of positive-norm states can be made. In section 5 we obtain more
general realisations, and make concluding remarks.

\bigskip
\noindent
{\bf 2. Miura Transformation}
\bigskip
      A realisation for the $N=2$ super-$W_{n+1}$ algebra is constructed by
the Miura transformation based on the $A(n, n-1)$ super algebra. It is most
conveniently expressed in terms of $N=1$ superfields [10], and is given by
the differential operator
$$
\eqalign{
{\cal M}_n&=\Big(\prod_{j=1}^n
[(\alpha_0 D+D \Phi_j-\chi_{n-j}^{(n)})(\alpha_0 D+D
\Phi_{j}-\chi_{n+1-j}^{(n)})]\Big)(\alpha_0 D-\chi_n^{(n)})\cr
&=(\a D)^{2n+1}+\sum_{\ell=2}^{2n+1} U_\ell^{(n)} (\a D)^{2n+1-\ell}
\ ,\cr}\eqno(2.1)
$$
where
$$
\eqalign{
\chi_0^{(n)}&=0\cr
\chi_j^{(n)}&=\sum_{k=n-j+1}^n D {\bar\Phi}_{k}(z, \theta)\ ,\cr}\eqno(2.2)
$$
and the $N=1$ superfields $\Phi_i(z,\theta)$ and ${\bar\Phi}_i(z,\theta)$
are given in terms of components by
$$
\eqalign{
\Phi_i(z,\theta)&=\phi_i(z)+\theta\psi_i(z)\ ,\cr
{\bar\Phi_i}(z,\theta)&={\bar\phi}_i(z)+\theta{\bar\psi}_i(z)
\ .\cr}\eqno(2.3)
$$
Here $D$ is given by
$$
D={\partial\over{\partial\theta}}+\theta\partial,\eqno(2.4)
$$
where $z$ and $\theta$ are the bosonic and fermionic coordinates of a
2-dimensional superspace, $\partial\equiv\partial/\partial z$  and the
derivative $D$ satisfies $D^2=\partial$. In equation (2.1), $U_\ell^{(n)}$
is a current of super-conformal spin $\ell/2$ in the super-$W_{n+1}$
algebra. In component language, it can be written as
$$
\eqalign{
\a^{-k}U_{2k}^{(n)}(z,\theta)&=B_k(z)+\theta F_{k+\ft12}(z)\ ,\cr
\a^{-k} U_{2k+1}^{(n)}(z,\theta)&=\widetilde F_{k+\ft12}(z)+\theta \widetilde
B_{k+1} (z)\ .\cr}\eqno(2.5)
$$
The components $B$ and $F$ denote the bosonic and fermionic currents with
spins indicated by their indices.

      By convention, we shall always order products such as the one in (2.1)
in decreasing order of $j$, {\it i.e.}\ the largest-$j$ factor sits at the
left. As a consequence, we can rewrite (2.1) as
$$
\eqalign{
{\cal M}_n&=(\a D+D\Phi_n)(\a D+D\Phi_n-D\bar\Phi_n) \cr
&\times \Big( \prod_{j=1}^{n-1} (\a D+D\Phi_j-\chi_{n-j-1}^{(n-1)}-
D\bar\Phi_n) (\a D+D\Phi_j-\chi_{n-j}^{(n-1)}-D\bar\Phi_n)\Big )\cr
&\times (\a D-\chi_{n-1}^{(n-1)}- D\bar\Phi_n) \ ,\cr}\eqno(2.6)
$$
where we have used the recursion relation of $\chi_i^{(n)}$:
$$
\chi_i^{(n)}=\chi_{i-1}^{(n-1)} +D\bar\Phi_n  \ ,\eqno(2.7)
$$
which follows immediately from (2.2). Since $\del +(\del f)=e^{-f}\del e^f$,
one can re-express equation (2.6) in the recursive form
$$
{\cal M}_n=(\a D+D\Phi_n)(\a D+D\Phi_n-D\bar\Phi_n) e^{\bar\Phi_n/\a}
{\cal M}_{n-1} e^{-\bar\Phi_n/\a}\ .\eqno(2.8)
$$
Using (2.1) with $n\to n-1$, we can therefore write the currents of
super-$W_{n+1}$ in terms of those of super-$W_n$ and an additional complex
pair of superfields $(\Phi_n, \bar\Phi_n)$. Explicitly, we find
$$
\eqalign{
U^{(n)}_{2j}&=\sum^{j}_{i=0}{n+i-j \choose i} \Big [ U^{(n-1)}_{2j-2i} - D
\bar\Phi_n U^{(n-1)}_{2j-2i-1} + \big ( - D\Phi_n D\bar\Phi_n +
\a^2 \partial \Phi_n \big ) U^{(n-1)}_{2j-2i-2} \cr
&+\a D\bar\Phi_n (D U^{(n-1)}_{2j-2i-2} )+\a^2 \partial U^{(n-1)}_{2j-2i-2}
\Big ] P_i(\bar\Phi_n)\ ,\cr
U^{(n)}_{2j+1}&=\sum^j_{i=0}{n+i-j \choose i} \Big [(\a D+D\Phi_n)(\a D + D
\Phi_n -D \bar\Phi_n) \Big( (\a D P_i(\bar\Phi_n) - P_i(\bar\Phi_n)
D\bar\Phi_n )U^{(n-1)}_{2j-2i-2} \cr
&+ P_i(\bar\Phi_n) U^{(n-1)}_{2j-2i-1} \Big)
+P_i(\bar\Phi_n) U^{(n-1)}_{2j-2i+1} + \a D P_i(\bar\Phi_n)
U^{(n-1)}_{2j-2i}\cr
&-P_{i+1}(\bar\Phi_n) U^{(n-1)}_{2j-2i-1} -
\a D P_{i+1}(\bar\Phi_n) U^{(n-1)}_{2j-2i-2}\Big ] \ ,\cr}\eqno(2.9)
$$
where the currents $U^{(n-1)}_j$ with $j < 0$ or $j=1$ are defined to be
zero, and $U^{(n-1)}_0=1$.  In (2.9) we have defined $P_i(\bar\Phi_n)$,
which is a differential polynomial in $\bar\Phi_n$, by
$$
P_i(\bar\Phi_n)\equiv e^{\bar\Phi_n/\a}\Big ( (\a^2\partial)^i
e^{-\bar\Phi_n/\a}\Big )\ .\eqno(2.10)
$$

      Equation (2.9) gives a realisation of the super-$W_{n+1}$ currents in
terms of those for super-$W_n$, together with an additional pair of complex
superfields $(\Phi_n, \bar\Phi_n)$.  Applying this reduction recursively,
one obtains a realisation of the super-$W_{n+1}$ algebra in terms of
$(\Phi_1,\bar\Phi_1)$, which appear in the currents only {\it via} their
super energy-momentum tensor, together with $(n-1)$ pairs of complex
superfields $(\Phi_2,\bar\Phi_2,\ldots,\Phi_n,\bar\Phi_n)$. Since
$(\Phi_1,\bar\Phi_1)$ commute with the other superfields, their super
energy-momentum tensor may be replaced by an arbitrary one that has the same
``exterior'' characteristics: {\it i.e.}\ commuting with
$(\Phi_2,\bar\Phi_2, \ldots,\Phi_n,\bar\Phi_n)$ and having the same central
charge.  This construction was given for the super-$W_3$ algebra in [1],
where both $N=2$ super currents and their components of the algebra were
given explicitly.

      The recursion formulae given in (2.9) are complicated in general;
fortunately, as we shall see later, it is unnecessary to solve the explicit
forms of the higher-spin super currents.  The lower-spin super currents can
be easily obtained from (2.9)
$$
\eqalign{
U^{(n)}_2&=U^{(n-1)}_2-D\Phi_n D\bar\Phi_n + \a \partial\Phi_n -n\a
\partial\bar\Phi_n \cr
&=\sum^n_{j=1} \Big (- D\Phi_j D\bar\Phi_j + \a\partial\Phi_j -j\a \partial
\bar\Phi_j \Big ) \ ,\cr
U^{(n)}_3&=U^{(n-1)}_3-\partial\Phi_n D\bar\Phi_n -n\a\partial D\bar\Phi_n
\cr
&=\sum^n_{j=1} \Big (- \partial\Phi_j D\bar\Phi_j +j\a \partial D \bar
\Phi_j \big) \ .\cr}\eqno(2.11)
$$
Introducing a second anticommuting coordinate $\tilde\theta$, one can
assemble the the currents $U^{(n)}_2$ and $U^{(n)}_3$ into the $N=2$ super
energy-momentum tensor $T(z, \theta,\tilde\theta)$ of the super-$W_{n+1}$
algebra
$$
T(z,\theta,\tilde\theta)=\ft14 U^{(n)}_2(z,\theta) +\ft14 \tilde\theta
\Big( 2U^{(n)}_3(z,\theta)-DU^{(n)}_2(z,\theta) \Big)\ ,\eqno(2.12)
$$
which can be expanded in components as
$$
T(z,\theta,\tilde\theta)=\ft12 J(z) + \ft12 \theta G_{\theta}(z) -\ft12
\tilde\theta
G_{\tilde\theta}(z) +\tilde\theta \theta T(z) \ .\eqno(2.13)
$$
The component currents $(J, G_{\theta}, G_{\tilde\theta}, T)$ have
conformal spins $(1,\ft32,\ft32,2)$ with respect to the energy-momentum
tensor $T(z)$. The explicit forms of these component currents are given by
$$
\eqalign{
J(z)&=\sum^n_{j=1} \big( -\psi_j\bar\psi_j +\a \partial\phi_j -j\a\partial
\bar\phi_j \big)\ ,\cr
T(z)&=\sum^n_{j=1} \big( \ft12\psi_j\partial\bar\psi_j -\ft12\partial\psi_j
\bar\psi_j -\partial\phi_j\partial\bar\phi_j -\ft12\a\partial^2\phi_j
-\ft12 j\a\partial^2\bar\phi_j \big)\ ,\cr
G(z)&\equiv \ft{1}{\sqrt{2}} \big( G_{\tilde\theta} + G_{\theta}
\big)=\sum^{n}_{j=1}\sqrt{2}\big(\partial\bar\phi_j\psi_j+\a\partial\psi_j
\big) \ ,\cr
\bar G(z)&\equiv \ft{1}{\sqrt{2}} \big( G_{\tilde\theta} - G_{\theta}
\big)=\sum^{n}_{j=1}\sqrt{2}\big(\partial\phi_j\bar\psi_j+j\a
\partial \bar\psi_j \big) \ ,\cr}\eqno(2.14)
$$
where the free fields $(\phi_j,\bar\phi_j)$ and $(\psi_j,\bar\psi_j)$ are
the components of the superfields defined in (2.3).  It follows from (2.14)
that the super energy-momentum tensor $T(z,\theta,\tilde\theta)$ given in
(2.12) generates the super-Virasoro algebra with central charge
$$
c_n=3n\Big( 1+(n+1)\a^2 \Big)\ .\eqno(2.15)
$$

    The pair of complex superfields $(\Phi_1,\bar\Phi_1)$ appears in (2.9)
only {\it via} $U^{(1)}_2$ and $U^{(1)}_3$, {\it i.e.} their super
energy-momentum tensor $T^{\rm eff}(z,\theta,\tilde\theta)=\ft14
U^{(1)}_2(z,\theta) +\ft14 \tilde\theta \big(
2U^{(1)}_3(z,\theta)-DU^{(1)}_2(z,\theta) \big)$. The components of this
super energy-momentum tensor are given by
$$
\eqalign{
J^{\rm eff}(z)&=-\psi_1\bar\psi_1 +\a\partial\phi_1-
\a\partial\bar\phi_1 \ ,\cr
T^{\rm eff}(z)&=\ft12\psi_1\partial\bar\psi_1-\ft12\partial\psi_1\bar \psi_1
-\partial\phi_1\partial\bar\phi_1 - \ft12\a \partial^2\phi_1 -\ft12\a
\partial^2 \bar\phi_1 \ ,\cr
G^{\rm eff}& =\sqrt{2}\big(\partial \bar\phi_1 \psi_1 + \a \partial \psi_1
\big ) \ ,\cr
\bar G^{\rm eff}&=\sqrt{2}\big(
\partial\phi_1\bar\psi_1+\a\partial\bar\psi_1\big)\ . }\eqno(2.16)
$$
Thus the super energy-momentum tensor $T^{\rm eff}(z,\theta,\tilde\theta)$
generates the super-Virasoro algebra with central charge $c^{\rm eff}$
given by
$$
\eqalign{
c^{\rm eff}&=3+6\a^2 \cr
&={2(c_n-6n) \over n(n+1)}+6 -\big (3-{6 \over n+1}\big)\ .
\cr}\eqno(2.17)
$$
The contribution from $(\Phi_1,\bar\Phi_1)$ can then be replaced by an
arbitrary super energy-momentum tensor with the same central charge given in
(2.17).

     In section (3), we shall prove that the solutions to the physical-state
conditions for the tachyonic states of the super-$W_{n+1}$ string form a
representation of the Weyl group for the bosonic subalgebra $A_n\!\oplus\!
A_{n+1}$ of the
super algebra $A(n,n-1)$.  For
this purpose we shall now describe the root space for this super algebra.
The Miura transformation (2.1) is a realisation in a particular basis of the
general expression
$$
{\cal M}_n=\prod^{2n}_{j=0} \big (\a D +  {\bf\vec H}^{(n)}_j\cdot D
{\bf \vec\Phi}^{(n)\dagger} \big )\ ,\eqno(2.18)
$$
where ${\bf \vec
\Phi}^{(n)\dagger}\equiv(\bar\Phi_1,\Phi_1;\,\bar\Phi_2,\Phi_2;\ldots;\,
\bar\Phi_n,\Phi_n)$ which interchanges the components in a
particular way from the vector superfield $
{\bf\vec\Phi}^{(n)}\equiv(\Phi_1,\bar\Phi_1;\, \Phi_2,\bar\Phi_2;\ldots;
\Phi_n,\bar\Phi_n)$.  In the conventions of this paper, for any
$2k$-component vector ${\bf\vec A}=(a_1,\bar a_1;\, a_2,\bar a_2;\ldots;
a_k,\bar a_k)$, ${\bf\vec A}^{\dagger}$ is defined by ${\bf\vec
A}^{\dagger}=(\bar a_1,a_1;\, \bar a_2, a_2;\ldots; \bar a_k,a_k)$. Note
that
$$
{\bf\vec A} \cdot {\bf\vec B}^{\dagger}={\bf\vec A}^{\dagger} \cdot
{\bf\vec B} \ .\eqno(2.19)
$$
The ${\bf\vec H}^{(n)}_j$ in (2.18) are $2n$-component vectors.
The Miura transformation (2.18) gives a realisation of super $W_{n+1}$
provided  that they satisfy
$$
\eqalign{
{\bf\vec H}^{(n)}_i\cdot {\bf\vec H}^{(n)\dagger}_j&=(-1)^j
\delta_{ij}-1 \ ,\cr
\sum^{2n}_{j=0} (-1)^j {\bf\vec H}^{(n)}_j&=0\ .\cr}\eqno(2.20)
$$
The Miura transformation (2.1) corresponds to the choice
$$
\eqalign{
{\bf\vec H}^{(n)}_{0}&=\Big(-1,0;\,-1,0;\,\ldots;\,-1,0\Big)\ ,\cr
{\bf\vec H}^{(n)}_{2k}&=\Big (\underbrace{0,0;\,\ldots;\,0,0}_{2(k-1)};\,
0,1;\,\underbrace{-1,0;\,-1,0;\ldots;\,-1,0}_{2(n-k)} \Big )\ ,
\qquad 1 \le k \le n \ ,\cr
{\bf\vec H}^{(n)}_{2k+1}&=\Big (\underbrace{0,0;\ldots,0,0}_{2k};\,
-1,1;\,\underbrace{-1,0;\,-1,0;\, \ldots;\,-1,0}_{2(n-k-1)}
\Big )\ , \qquad 0 \le k \le n-1 \ .\cr}
\eqno(2.21)
$$

The vectors ${\bf\vec H}^{(n)}_j$ for $0 \le j \le (2n-1)$ are the weights
of the defining representation of $A(n,n-1)$.
The simple roots ${\bf\vec e}^{(n)}_j$ of $A(n,n-1)$ can be given
in terms of these weights by
$$
{\bf\vec e}^{(n)}_j=(-1)^j\big({\bf\vec H}^{(n)}_j
-{\bf\vec H}^{(n)}_{j+2}\big)\ ,
\qquad 0\le j\le 2n-1\ ,\eqno(2.22)
$$
where ${\bf\vec H}_{2n+1}^{(n)}$ is defined to be zero. The vectors ${\bf
\vec e}_{2p}^{(n)}$, $0\le p\le n-1$,  are the simple roots for $A_n$, and
${\bf \vec e}_{2p+1}^{(n)}$, $0\le p\le n-2$, are the simple roots for
$A_{n-1}$ in the bosonic $A_n\!\oplus\! A_{n-1}$ subalgebra of $A(n,n-1)$.
For later purposes we also introduce the Weyl vector
$\vec\bfrho^{\phantom{0}\! (n)}$, given
by
$$
\eqalign{
\vec\bfrho^{\phantom{o}\! (n)}&=\ft12\sum^{2n-1}_{j=0} (-1)^j
\Big( 2n-j\Big) {\bf\vec H}^{(n)}_j \cr
&=\ft12\sum_{p=0}^{n-1} (p+1)(n-p){\bf\vec e}_{2p}^{(n)}
+\ft12\sum_{p=0}^{n-2} (p+1)(n-1-p){\bf\vec e}_{2p+1}^{(n)}
\cr
&=\vec\bfrho\, (A_n)+\vec\bfrho\, (A_{n-1})\ ,\cr}\eqno(2.23)
$$ where $\vec\bfrho\, (A_n)$ and $\vec\bfrho\, (A_{n-1})$ are the Weyl
vectors for the $A_n$ and $A_{n-1}$ factors in the $A_n\!\oplus\! A_{n-1}$
bosonic subalgebra. In the basis of (2.21), the Weyl vector is
$$
{\vec\bfrho}^{\phantom{0}\! (n)}=\Big ( \underbrace{-\ft12,-\ft12;\,-1,
-\ft12;\, -\ft32,-\ft12;\,\ldots;-\ft{n}{2}, -\ft12}_{2n}
\Big ) \ . \eqno(2.24)
$$
In vector notation, one can rewrite the energy-momentum tensor in (2.14)
in a nicer form
$$
T(z)=\ft12 \vec\bfpsi^{(n)}\cdot\partial\vec\bfpsi^{(n)\dagger}
-\ft12\partial\vec\bfphi^{(n)}\cdot\partial\vec\bfphi^{(n)\dagger}
+\a\vec \bfrho^{\phantom{0}\! (n)} \cdot\partial^2
\vec \bfphi^{(n)\dagger}\ ,\eqno(2.25)
$$
where
$\vec\bfpsi^{(n)}\equiv(\psi_1,\bar\psi_1;\,\psi_2,\bar\psi_2;\,\ldots;
\,\psi_n,\bar\psi_n)$ and $\vec\bfphi^{(n)}\equiv(\phi_1,\bar\phi_1;\,
\phi_2,\bar\phi_2;\,\ldots;\, \phi_n,\bar\phi_n)$.

        Note that the closure of the super-$W_{n+1}$ algebra requires only
that the vectors ${\bf\vec H}^{(n)}_j$ in (2.18) satisfy conditions (2.20).
There exist many solutions to (2.20).  If one only considers the realisation
in terms of $n$ pairs of complex superfields, all these solutions are
equivalent: an $SO(n,n)$ transformation of ${\bf\vec \Phi}$ will map one
solution to another.  However, the particular solution given in (2.21) has
the nice property that one can express the currents of the super-$W_{n+1}$
algebra in terms of those of super-$W_n$ with an additional pair of complex
superfields. This property is essential to construct the string theory since
it implies that one eventually obtains a realisation of the super-$W_{n+1}$
algebra in terms of an arbitrary super energy-momentum tensor together with
$(n-1)$ additional pairs of complex superfields.  We shall see later that
the scalar fields in these $(n-1)$ additional pairs of complex superfields
do not describe physically-observable coordinates.  It is essential that
this super energy-momentum tensor can be arbitrary; and therefore, if it
contains $D$ pairs of complex superfields $(\Phi_{\mu},\bar\Phi_{\mu})$ one
obtains a string theory effectively describing a real $2D$-dimensional
target spacetime.

\bigskip
\bigskip
\noindent
{\bf 3. The Physical Spectrum of the Super-$W_{n+1}$ String}
\bigskip
\bigskip

      In this section, we shall construct the complete spectrum of physical
states of the super-$W_{n+1}$ string; we shall suppress the superscript
index $(n)$ when there is no ambiguity. As we have mentioned in the
introduction section, the anomaly-free condition of the string theory
requires that the central charge take its critical value. In our case, that
means the central charge given in (2.15) should take the value given in
(1.7$a$). This implies that the background charge parameter $\a$ takes its
critical value $\a^*$ given by
$$
(\a^*)^2={1\over{n+1}}\ .\eqno(3.1)
$$

    In order to construct a string theory that allows transverse
excitations, it is necessary to add extra coordinates as we indicated at the
end of section 2. Since these extra coordinates enter only {\it via}\  the
effective super energy-momentum tensor whose components are given in (2.16),
it is sufficient for the purposes of determining the physical spectrum to
work with the basic Miura realisation without additional superfields.

\bigskip
\noindent
{\it 3.1 The physical-state conditions}
\bigskip

The physical-state conditions are determined by the requirement that the
product of a physical state and the ghost vacuum should be BRST invariant.
This leads to the following conditions
$$\eqalign{
(B_k)_m\ket{\rm phys}&=0,\qquad
(\widetilde B_{k+1})_m\ket{\rm phys}=0 ,\ m\ge 1\cr
(F_{k+1/2})_r\ket{\rm phys}&=0,\qquad
({\widetilde F}_{k+1/2})_r\ket{\rm phys}=0 ,\ r\ge \ft12\cr
(B_k)_0\ket{\rm phys}&=\omega_k\ket{\rm phys},\qquad
(\widetilde B_{k+1})_0\ket{\rm phys}=\widetilde\omega_{k+1}
\ket{\rm phys} \ ,\cr}\eqno(3.2)
$$
where $k=1,2,3,\ldots, n$, and the component currents are defined in (2.5).
The constants $\omega_k$ and $\widetilde\omega_{k+1}$ are the intercepts for
the bosonic currents, which are in principle determined by the BRST
invariance condition stated above.  Owing to the complexities of the
non-linear super-$W_{n+1}$ algebra, it is difficult in general to determine
the intercepts by this method.  However, since the super-Virasoro algebra
forms a linear subalgebra, one can straightforwardly read off from the
structure of the ghost vacuum which was described in section 1 that the
intercepts for both the spin-1 current and the spin-2 energy-momentum tensor
are zero [1].  For the remaining intercepts, we resort to a generalisation
of an argument first introduced in [6], and developed in [7].  It was
conjectured in [6] that for any $W$ string there is always a particular
tachyonic physical state created by the action of the operator
$\exp(\lambda\a^*\vec\rho\cdot\vec\varphi)$ on the vacuum, for some constant
$\lambda$, where $\vec\rho$ is the Weyl vector of the underlying Lie
algebra. The constant $\lambda$ can be determined from the known value of
the spin-2 intercept. In [7] it was demonstrated  for the $W_3$, $W_4$ and
$W_5$ strings that the requirement of unitarity singles out intercept values
that correspond to these particular states.  It is very plausible that an
operator of this form, known as the ``cosmological operator,'' will always
create a physical state, in any $W$-string theory.  Assuming this to be the
case for the super-$W_{n+1}$ string, {\it i.e.}\ that
$$
e^{\lambda\a^*\vec\bfrho\cdot\vec\bfphi^\dagger} \eqno(3.3)
$$
satisfies the physical condition (3.2), it follows from (1.7b) that
$\lambda$ must satisfy
$$
\lambda (\lambda-2)=0\ .\eqno(3.4)
$$
Taking the root $\lambda=0$, it is clear from (3.2) all the intercepts
$\omega_k$ and $\widetilde\omega_{k+1}$ are zero. (In fact in any $W$ string
theory where there is a physical state of cosmological type, and the $L_0$
intercept is zero, it will always be the case that all other intercepts
vanish too.) Note that if the intercepts all vanish in one basis, then they
vanish in all bases. In particular, for example, they will vanish in the
primary basis. A different method, based on the analysis of null states, was
used in [1] to show that all the intercepts in the super-$W_3$ are zero.
This provides some evidence in support of the cosmological conjecture.

\bigskip\bigskip
\noindent{\it 3.2 Tachyonic Operators and the Weyl Group  }
\bigskip\bigskip

     Tachyonic, {\it i.e.}\ level-0, states are built from physical
operators involving only the bosonic fields complex $\vec\bfphi$, namely
$$
e^{\vec\bfbeta\cdot \vec\bfphi^\dagger}.\eqno(3.5)
$$
Of the physical-state conditions (3.2), the fermionic ones are therefore
empty; the bosonic conditions for positive Laurent modes are automatically
satisfied, and the only non-trivial conditions come from the intercept
equations.  These give a set of polynomial equations for $\vec\bfbeta$.  The
number of equations, which are non-degenerate, is equal to the number of
components of $\vec\bfbeta$, and so the solutions are discrete.  The number
of such solutions is given by the product of the degrees of the polynomials,
{\it i.e.}\ by the product of the spins of the bosonic currents.  Thus we
have $n!(n+1)!$ discrete solutions for $\vec\bfbeta$.  As we shall now show,
the polynomial equations are invariant under the action of a discrete
symmetry group of dimension $n!(n+1)!$ which acts transitively on the
solutions.  This is in fact the Weyl group of $A(n,n-1)$. Thus we can obtain
{\it all} tachyonic solutions by acting with the Weyl group on any one
solution.  Since the cosmological solution is assumed to be a physical
state, all the other tachyonic physical states can therefore be obtained by
acting on it with the Weyl group.

     Since the tachyonic physical states do not involve fermions, we may set
the fermion fields in the currents to zero when calculating the momentum
polynomials coming from the intercept conditions in (3.2). This implies we
can just keep $\theta\del\vec\bfphi$ in $D\vec{\bf \Phi}$. The eigenvalues
of the zero modes of the bosonic currents acting on the tachyonic state can
be read off from the highest order pole of the OPE between the bosonic
currents and the tachyonic physical operator. Since
$$
\del\vec\bfphi(z)\ e^{\vec\bfbeta\cdot\vec\bfphi^\dagger}(w)\sim
-{{\vec\bfbeta}\over{z-w}}e^{\vec\bfbeta\cdot\vec\bfphi^\dagger}(w)\ ,
\eqno(3.6)
$$
it follows that one can obtain the eigenvalues of the zero modes of the
bosonic currents directly from Miura transformation (2.1) and (2.18) by the
replacing $D\vec{\bf\Phi}\to -\ft{\theta\vec\bfbeta}{z}$, {\it i.e.}\
$$
\prod_{j=0}^{2n}\big(\a D-\theta\vec {\bf H}_j\cdot{\vec\bfbeta^\dagger\over
z}\big)
=(\a D)^{2n+1}+\sum_{p=1}^{n} {\a^p{b_p(\vec\bfbeta)}\over {z^p}}(\a
D)^{2n-2p+1} +\theta\sum_{q=1}^n {\a^q{{\widetilde b}_{q+1}(\vec\bfbeta)}\over
{z^{q+1}}} (\a D)^{2n-2q}
\ ,\eqno(3.7)
$$
where $b_p(\vec\bfbeta)$ and ${\widetilde b}_q(\vec\bfbeta)$, which are
polynomials in $\vec\bfbeta$, are the eigenvalues of the zero modes of the
bosonic currents $B_p$ and ${\widetilde B}_q$. By acting with (3.7) on $z^j$
and $\theta z^j$, for $1\le j\le n$, we have
$$
\eqalignno{
\a^{2n+1}\prod_{p=0}^n (j-\ft12 n-\ft12 {\bf \vec
H}_{2p}\cdot\vec\bfgamma^\dagger) &=\sum_{s=0}^j {j!\over{(j-s)!}}
\big(b_{n+1-s}(\vec\bfbeta)+{\widetilde b}_{n+1-s}(\vec\bfbeta)\big)
\a^{n+s}\ &(3.8a)\cr
\a^{2n+1}\prod_{p=1}^n \big(j-\ft12 (n+1)-\ft12{\bf \vec H}_{2p-1}
\cdot\vec\bfgamma^\dagger\big)
&=\sum_{s=0}^j {j!\over{(j-s)!}} b_{n-s}(\vec\bfbeta)\a^{n+s+1} \ ,&(3.8b)\cr}
$$
where $b_{n+1}(\vec \bfbeta)$ and $\widetilde b_1(\vec\bfbeta)$ are both
equal to zero. The vector $\vec\bfgamma$ in (3.8a) and (3.8b) is the shifted
momentum defined by
$$
\vec\bfbeta=\a(\ft12\vec\bfgamma+\vec\bfrho\,)\ .\eqno(3.9)
$$

     It is clear that the equations (3.8$a$) and (3.8$b$) are invariant
under independent permutations of the ${\bf \vec H}_{2p}$'s and of the ${\bf
\vec H}_{2p-1}$'s. Thus the eigenvalues $b_p$ and ${\widetilde b}_q$ are
invariant under a discrete symmetry of order $n!(n+1)!$. In fact, this
symmetry group is just the Weyl group of $A_n\!\oplus\! A_{n-1}$. To see
this, we note that under a Weyl reflection $S_j$ corresponding to the simple
root ${\bf \vec e}_j$, the simple root ${\bf \vec e}_i$ transforms as
$$
\eqalign{
S_j({\bf \vec e}_i)&= {\bf \vec e}_i
-(-1)^j({\bf \vec e}_i\cdot{\bf \vec e}_j^\dagger){\bf \vec e}_j\cr
&={\bf \vec e}_i-(2\delta_{ij}-\delta_{i,j+2}-\delta_{i+2,j})
{\bf \vec e}_j\ .\cr}
\eqno(3.10)
$$
{}From (2.22) it follows that $S_j$ acting on the ${\bf \vec H}_p$'s only
interchanges ${\bf \vec H}_j$ with ${\bf \vec H}_{j+2}$, leaving all others
unchanged. Since the scalar product is invariant,  a Weyl reflection of
$\vec\gamma$
$$
\vec\bfgamma\longrightarrow S_j(\vec\bfgamma)=\vec\bfgamma-(-1)^j
(\vec\bfgamma\cdot
{\bf \vec e}_j^\dagger){\bf \vec e}_j
\eqno(3.11)
$$
therefore leaves the left-hand sides of (3.8$a$) and (3.8$b$) invariant. The
dimension of the Weyl group of $A_n\!\oplus\! A_{n-1}$ is $n!(n+1)!$. All
the elements can be generated from the elements $S_j$ corresponding to the
simple roots. For example, they can be generated by choosing one entry from
each column of the following
$$
\eqalign{
\left\lgroup\matrix{1\cr S_1\cr}\right\rgroup &\otimes
\left\lgroup\matrix{1\cr S_3\cr S_3 S_1\cr}\right\rgroup\otimes
\left\lgroup\matrix{1\cr S_5\cr S_5 S_3\cr S_5 S_3
S_1\cr}\right\rgroup\otimes \cdots\otimes
\left\lgroup\matrix{1\cr S_{2n-3}\cr S_{2n-3} S_{2n-5}\cr\vdots\cr
S_{2n-3}S_{2n-5}\cdots S_1\cr}\right\rgroup\cr
\otimes\left\lgroup\matrix{1\cr S_0\cr}\right\rgroup&\otimes
\left\lgroup\matrix{1\cr S_2\cr S_2 S_0\cr}\right\rgroup\otimes
\left\lgroup\matrix{1\cr S_4\cr S_4 S_2\cr S_4 S_2
S_0\cr}\right\rgroup\otimes \cdots\otimes
\left\lgroup\matrix{1\cr S_{2n-2}\cr S_{2n-2} S_{2n-4}\cr\vdots\cr
S_{2n-2}S_{2n-4}\cdots S_0\cr}\right\rgroup\ .\cr}
\eqno(3.12)
$$
Thus the polynomials $b_p$ and ${\widetilde b}_q$ are invariant under the
action of the entire Weyl group on the shifted momentum $\vec\bfgamma$.

     The upshot of the above discussion is that the Weyl group maps physical
states into physical states. By supposition, the cosmological operator (3.3)
with $\lambda=0$ gives a physical state. From (3.9), it follows that the
shifted momentum for the cosmological solution is
$$
\vec\bfgamma^{\rm cosmo}=-2\vec\bfrho\ .\eqno(3.13)
$$
The Weyl group acts without fixed points on the Weyl vector $\vec\bfrho$, and
so we generate all the $n!(n+1)!$ distinct tachyonic physical states.

\bigskip
\noindent
{\it 3.3  The Spectrum of Physical states}
\bigskip

     We have seen in section 3.2 that in the Miura realisation for
super-$W_{n+1}$ in terms of $n$ pairs of complex superfields, {\it all}\ the
momentum components of tachyonic states are fixed to a set of discrete
values by the physical-state conditions. Clearly the scalars conjugate to
these momentum components are not at this stage physically-observable
coordinates. When we add additional superfields to the effective super
energy-momentum tensor, the momentum components $\beta_2, \beta_3, \ldots,
\beta_n, \bar\beta_2, \bar\beta_3, \ldots, \bar\beta_n$ remain frozen to the
same values, whilst $\beta_1$ and $\bar\beta_1$, together with momenta
conjugate to the extra coordinates, can now take continuous values subject
to constraints imposed by the effective intercepts for $L_0^{\rm eff}$ and
$J_0^{\rm eff}$.

     We consider the $T^{\rm eff}(z,\theta,\tilde\theta)$ with
$D$ pairs of complex fields $(\Phi_\mu,\bar\Phi_\mu)$,
$\mu=0,1,\ldots,(D-1)$.  In component form, the currents of the effective
super energy-momentum tensor given in (2.16) are modified to become
$$
\eqalign{
J^{\rm eff}(z)&=-\psi_\mu\bar\psi^\mu+Q_\mu \partial
\phi^\mu-Q_\mu\partial\bar\phi^\mu \ ,\cr
T^{\rm eff}(z)&=\ft12\psi_\mu\partial\bar\psi^\mu-\ft12\partial\psi_\mu
\bar \psi^\mu-\partial\phi_\mu\partial\bar\phi^\mu -
\ft12 Q_\mu \partial^2\phi^\mu -\ft12 Q_\mu \partial^2
\bar\phi^\mu \ ,\cr
G^{\rm eff}& =\sqrt{2}\big(\partial \bar\phi_\mu \psi^\mu+ Q_\mu
\partial \psi^\mu\big ) \ ,\cr
\bar G^{\rm eff}&=\sqrt{2}\big(\partial\phi_\mu\bar\psi^\mu
+Q_\mu\partial\bar\psi^\mu\big)\ . }\eqno(3.14)
$$
The central charge for $T^{\rm eff}$ must be given by
$$
c^{\rm eff}=3+6(\a^*)^2=3+{6\over{n+1}}\ .\eqno(3.15)
$$
This implies that the background charge vector $Q_\mu$ must satisfy
$$
\eqalign{
Q_\mu Q^\mu&=\a^{*2}+\ft12 (1-D)\cr
&={1 \over n+1} +\ft12 (1-D)\cr}\ .\eqno(3.16)
$$
Since the values of the intercepts $\omega_1^{\rm eff}$ and $\omega_2^{\rm
eff}$ for  $J^{\rm eff}_0$ and $L^{\rm eff}_0$ are independent of whether or
not additional superfields are included in the effective super
energy-momentum tensor, it follows from (2.16) that they are given by
substituting the discrete solutions for $\beta_1$ and $\bar\beta_1$ found in
section 3.2 into
$$
\eqalignno{
\omega^{\rm eff}_1&=-\a^*\beta_1+\a^*\bar\beta_1
={1\over 2(n+1)}\big(\bar\gamma_1-\gamma_1\big) &(3.17a)\cr
\omega^{\rm eff}_2&=-\beta_1\bar\beta_1-\ft12\a^*\beta_1
-\ft12\a^*\bar\beta_1 =-{1\over4(n+1)}\big(\gamma_1\bar\gamma_1-1\big)
&(3.17b)\cr}
$$
{}From the discussion in section 3.2, we can find all the values of $\gamma_1$
and $\bar\gamma_1$ by acting on $\vec\bfgamma^{\rm cosmo}$ with the Weyl
group. We shall discuss these values below.

      Turning now to the higher level states, we have to divide them into
two categories. The first consists of states where the excitations lie
exclusively in the unfrozen directions $(\Phi_\mu,\bar\Phi_\mu)$. The second
category consists of states that include excitations in the frozen
directions. For reasons that we shall discuss later, all the physical states
in the second category seem to have zero norm and hence do not appear in the
physical spectrum. We shall therefore concentrate for now on physical states
in the first category. These can be written as
$$
\ket{\rm phys}=e^{\beta_2\bar\phi_2 +\cdots+\beta_n\bar\phi_n
+\bar\beta_2\phi_2+\cdots+\bar\beta_n\phi_n} \ket{\rm phys}_{\rm eff}
\ .\eqno(3.18)
$$
It is clear that (3.17) will satisfy the physical-state conditions (3.2)
provided that $\beta_2, \beta_3, \ldots,
\beta_n, \bar\beta_2, \bar\beta_3, \ldots, \bar\beta_n$ take their frozen
values found in section 3.2, and $\ket{\rm phys}_{\rm eff}$ satisfies the
physical-state conditions for the effective super energy-momentum tensor
$$\eqalign{
J^{\rm eff}_m\ket{\rm phys}_{\rm eff}&=0,\qquad
L_m^{\rm eff}\ket{\rm phys}_{\rm eff}=0 ,\ m\ge 1\cr
G^{\rm eff}_r\ket{\rm phys}_{\rm eff}&=0,\qquad
\bar G^{\rm eff}_r\ket{\rm phys}_{\rm eff}=0 ,\ r\ge \ft12\cr
J^{\rm eff}_0\ket{\rm phys}_{\rm eff}
&=\omega_1^{\rm eff}\ket{\rm phys}_{\rm eff},\qquad
L^{\rm eff}_0\ket{\rm phys}_{\rm eff}=\omega_2^{\rm eff}
\ket{\rm phys}_{\rm eff} \ ,\cr}\eqno(3.19)
$$
Thus the physical spectrum for the super-$W_{n+1}$ string is given by the
physical spectra for a set of effective super-Virasoro strings with a
non-standard central charge (3.15) and a set of non-standard intercepts
given by (3.17$a,b$).

     The complete physical spectrum is now in principle determined, since
$\omega^{\rm eff}_1$ and $\omega^{\rm eff}_2$ are given by (3.17$a,b$), and
$\gamma_1$ and $\bar\gamma_1$ are determined by acting with the Weyl group
on $\vec\bfgamma^{\rm cosmo}$ given by (3.13).  We have examined the
examples of super-$W\!A_n$ for $n=2,3,4,5,6$ explicitly, and found that in
all these cases the results for the allowed values of
$(\gamma_1,\bar\gamma_1)$ are as follows:  Each of $\gamma_1$ and
$\bar\gamma_1$ can take its values from the set of integers
$\pm1,\pm3,\pm5,\ldots,\pm(2n-1)$.  All possible combinations from these
values can occur provided that the following two conditions are satisfied:
$$
\eqalignno{
\gamma_1&\ne -\bar\gamma_1,&(3.20a)\cr
\gamma_1\bar\gamma_1&\le (n^2-1).&(3.20b)\cr}
$$
We expect that these results will apply also for the general case. As we
shall now show, equations (3.20$a,b$) give $n(3n-1)$ solutions for
$(\gamma_1,\bar\gamma_1)$.  For the case that $\gamma_1\bar\gamma_1 < 0$,
only (3.20$a$) imposes a non-trivial restriction, so we have $2n(n-1)$  such
solutions for $(\gamma_1,\bar\gamma_1)$. They may be written as
$$
\eqalign{
\gamma_1&=\pm (2r-1)\ ,\cr
\bar\gamma_1&=\mp (2s-1)\ ,\cr}\eqno(3.21)
$$
with $1\le r\ne s\le n$.
These give $n(n-1)$ positive
values for the effective spin-2 intercept:
$$
\omega_2^{\rm eff}={1\over 2(n+1)}(2rs-r-s+1),\eqno(3.22)
$$
Correspondingly, the effective spin-1 intercepts are
$$
\omega_1^{\rm eff}=\pm{r+s-1\over n+1}.\eqno(3.23)
$$
When $\gamma_1\bar\gamma_1 > 0$, only (3.20$b$) imposes a non-trivial
restriction, and we have $n(n+1)$ such solutions for
$(\gamma_1,\bar\gamma_1)$.  They can be expressed as
$$
\eqalign{
\gamma_1&=m\pm (\ell+1),\cr
\bar\gamma_1&=-m\pm (\ell+1),\cr}\eqno(3.24)
$$
where $\ell=0,1\ldots,(n-1)$, and $m=-\ell,-\ell+2,\ldots,\ell$.  These give
$\ft12n(n+1)$ negative values for the effective spin-2 intercept:
$$
\omega_2^{\rm eff}=-{\ell(\ell+2)-m^2\over 4(n+1)}.\eqno(3.25)
$$
the corresponding effective spin-1 intercept values are
$$
\omega_1^{\rm eff}=-{m\over n+1}.\eqno(3.26)
$$

     In the case of bosonic $W$ strings, it was found that there is a
relation with minimal models [6,7,8,9]. As we shall now show, a similar
relation emerges for the case of super $W$ strings.  First of all, we note
from (3.15) that the effective central charge for $T^{\rm eff}$ can be
written as
$$
c^{\rm eff}=6-\Big(3-{6\over n+1}\Big).\eqno(3.27)
$$
Here, 6 is the critical central charge for the usual $N=2$ super-Virasoro
string, and the term in parentheses is the central charge of the $(n-1)$'th
$N=2$ superconformal minimal model [12].  When the effective spin-2
intercept $\omega_2^{\rm eff}$ is negative, as given by (3.25), we observe
that $-\omega_2^{\rm eff}$ and $-\omega_1^{\rm eff}$ take values equal to
the dimensions and $U(1)$ charges of the above superconformal minimal model.
As for the bosonic $W$ strings, the significance of these observations is
not fully understood.

\bigskip\bigskip
\noindent
{\bf 4. The Issue of Unitarity}
\bigskip\bigskip

     Having obtained the spectrum of physical states for the super-$W\!A_n$
string, we now address the question of whether these states are unitary.  As
discussed in section 3.3, physical states involve excitations only in the
unfrozen directions, and thus take the form (3.18).  The effective physical
states $\ket{\rm phys}_{\rm eff}$ satisfy super-Virasoro-like physical-state
conditions (3.19), with a non-standard central charge (3.15), and
non-standard bosonic intercepts $\omega_1^{\rm eff}$ and $\omega_2^{\rm
eff}$ given by (3.23), (3.26) and (3.22), (3.25).  Note that for each value
of $\omega_2^{\rm eff}$, $\omega_1^{\rm eff}$ can take two values, with
equal magnitude but opposite sign.  In order to investigate the unitarity of
the physical states of the super-$W_n$ string, it suffices to investigate
the unitarity of the corresponding effective superstring theories.

     We shall take the effective super-Virasoro algebra to be realised by
the currents (3.14).  The background-charge vector $Q_\mu$ will be taken to
lie along the $\mu=0$ direction.  Thus (3.14) becomes
$$
\eqalign{
J^{\rm eff}(z)&=-\psi_\mu\bar\psi^\mu - Q \partial
\phi_0+Q\partial\bar\phi_0 \ ,\cr
T^{\rm eff}(z)&=\ft12\psi_\mu\partial\bar\psi^\mu-\ft12\partial\psi_\mu
\bar \psi^\mu-\partial\phi_\mu\partial\bar\phi^\mu +
\ft12 Q \partial^2\phi_0 +\ft12 Q \partial^2 \bar\phi_0 \ ,\cr
G^{\rm eff}& =\sqrt{2}\big(\partial \bar\phi_\mu \psi^\mu- Q
\partial \psi_0\big ) \ ,\cr
\bar G^{\rm eff}&=\sqrt{2}\big(\partial\phi_\mu\bar\psi^\mu
-Q\partial\bar\psi_0\big)\ , }\eqno(4.1)
$$
with $Q$ given by
$$
Q^2=\ft12(D-1)-{1\over n+1}.\eqno(4.2)
$$
Note that for $n \ge 2$, $Q$ cannot be zero for any choice of (integer) $D$,
and so the necessity for background charges cannot be avoided.

    Effective physical states can be constructed by acting on an
$SL(2,C)$-invariant vacuum $\big|0\big\rangle$ with ground-state operators
$P(z)$, {\it i.e.}\ $\ket{\rm phys}_{\rm eff}\equiv P(0)\ket{0}$.  The
ground-state operators take the form
$$
P(z)=R(z) e^{\beta\cdot\bar\phi+\bar\beta\cdot\phi}\ .\eqno(4.3)
$$
The operators $R(z)$ can be classified by their eigenvalues $q$ and $\ell$
under $J^{\rm eff}_0$ and $L^{\rm eff}_0$ respectively.  The eigenvalue $q$
measures the fermion charge of the operator $R(z)$; each $\psi_\mu$ in a
monomial in $R(z)$ contributes $+1$, each ${\bar\psi}_\mu$ contributes $-1$,
and $\del\phi_\mu$ and $\del\bar\phi_\mu$ contribute 0. The eigenvalue
$\ell$ measures the conformal dimension of the operator $R(z)$, {\it i.e.}\
the level number.

    At level $\ell=0$, $R$ is just the identity operator, with $q=0$, and
$P(z)$ is the ``tachyon" ground-state operator. At level $\ell=\ft12$, $R$
can be $\bar\xi_\mu\psi^\mu$, with $q=+1$; or $\xi_\mu\bar\psi^\mu$, with
$q=-1$. At level $\ell=1$, $q$ can be $-2,\ 0,\ +2$. In general, at level
$\ell$, $q$ can take the values
$$
q=-2\ell,\ -2\ell+2,\ \ldots,\ 2\ell-2,\ 2\ell\ .\eqno(4.4)
$$

     For effective physical states with level number $\ell$ and fermion
charge $q$, the $J^{\rm eff}_0$ and $L^{\rm eff}_0$ constraints in (3.19)
give
$$
\eqalignno{
J_0:\qquad  \omega_1^{\rm eff}&=q+Q (\beta_0-\bar\beta_0)\ ,&(4.5a)\cr
L_0:\qquad \omega_2^{\rm eff}&=\ell -\beta^\mu \bar\beta_\mu +\ft12 Q
(\beta_0+\bar\beta_0)
\ .&(4.5b)\cr}
$$
The hermiticity conditions for $L^{\rm eff}_0$ and $J^{\rm eff}_0$ imply
that [13]
$$
\eqalignno{
\beta_i^*&=-\bar\beta_i\ ;\qquad \qquad{\bar\beta}_i^*=
-\beta_i\ , &(4.6a)\cr
\beta_0^*&=-\bar\beta_0-Q\ ;\qquad {\bar\beta}_0^*=-\beta_0-Q\ ,
&(4.6b)\cr}
$$
where $^*$ denotes complex conjugation, and $i=1,2,\cdots,D-1$.

     The norm of an effective physical state of the form $\ket{\rm
phys}_{\rm eff}= P(0)\ket{0}$, with $P(0)$ given by (4.3), is of the form
${\cal N}(R)\ \big\langle p\ket{p}$, where ${\cal N}(R)$ is a function of
the scalar products of the polarisation tensors.  The term $ \big\langle
p\ket{p}$ gives the usual momentum-conservation delta function, and so the
requirement of non-negativity of the norm amounts to the requirement that
${\cal N}(R)$ be non-negative whenever $\big\langle p\ket{p} \ne 0$.
Actually, the norms of the physical states of the super-$W_n$ string will
also involve momentum-conservation delta functions in the frozen directions
as well. These will have the form [7,14]
$$
\prod_{i=2}^n\Big(\delta(\gamma_i+\bar\gamma_i +\gamma_i'+\bar\gamma_i')
\delta(\gamma_i-\bar\gamma_i +\gamma_i'-\bar\gamma_i')\Big)\ . \eqno(4.7)
$$
We know from the discussion of the Weyl group in section 3 that if $\vec
\bfgamma$ satisfies the physical-state conditions then so does
$-\vec\bfgamma$, so there always exist pairs of states for which (4.7) is
non-zero.  This pair of states have equal $\omega_2^{\rm eff}$ values and
opposite $\omega_1^{\rm eff}$ values.

     The momentum conservation for the effective superstring states can
never be satisfied when the fermion number is non-zero.  To see this, we
observe from (4.5$a$) that
$$
\beta_0-\bar\beta_0=Q^{-1}(\omega_1^{\rm eff}-q).\eqno(4.8)
$$
Thus the momentum-conservation delta function for the real direction
$\phi_0-\bar\phi_0$, {\it i.e.}\ $\delta(\beta_0-\bar\beta_0 +\beta_0'
-\bar\beta_0')$, can be satisfied only if the right-hand side of (4.8) for
the unprimed state cancels against the contribution for the primed state. If
$q=0$, this is easily achieved, since the $\omega_1^{\rm eff}$ values (3.23)
or (3.26) always come in equal and opposite pairs, so the frozen
$\beta_0-\bar\beta_0$ value for an unprimed state with given $\omega_1^{\rm
eff}$ can be cancelled by the frozen value for a primed state with the
opposite sign for $\omega_1^{\rm eff}$.  If $q\ne0$, the only possibility
would be if the ``displaced'' intercept $\omega_1^{\rm eff}-q$ in (4.8) for
the unprimed state happened to equal the negative of another displaced value
for the primed state.  For the case (3.26) this cannot happen, since there
$|\omega_1^{\rm eff}|<1$, whilst the fermion charge must take integer
values.  For the case (3.23) it could occur.  However, the corresponding
$\omega_2^{\rm eff}$ values for the two states would then be unequal, which
would mean, in the light of the discussion above, that the frozen-momentum
conservation delta functions (4.7) would be zero.  Thus the effective
physical states with fermion charge $q\ne 0$ have zero norm, and we need not
consider them further.

     The first non-trivial physical states that we need consider are thus at
level $\ell=1$, with fermion charge $q=0$.  They have the form (4.3) with
$$
R(z)=\varepsilon_{\mu\nu}\psi^\mu\bar\psi^\nu
+\bar\xi_\mu\del\phi^\mu+\xi_\mu\del\bar\phi^\mu\ .\eqno(4.9)
$$
In addition to the $J^{\rm eff}_0$ and $L^{\rm eff}_0$ constraints, we have
three other independent nontrivial constraints, coming from $J^{\rm eff}_1$,
$G^{\rm eff}_{1/2}$ and ${\bar G}^{\rm eff}_{1/2}$. They give,
respectively,
$$
\eqalignno{
\varepsilon^\mu{}_\mu&=Q(\xi_0-\bar\xi_0)\ ,&(4.10a)\cr
\bar\xi^\mu&=\beta_\nu^*\ \varepsilon^{\mu\nu}\ ,&(4.10b)\cr
\xi^\mu&=-{\bar\beta}_\nu^*\ \varepsilon^{\nu\mu}\ .&(4.10c)\cr}
$$
The norm $\cal N$ for these states is given by
$$
{\cal N}=\varepsilon^*_{\mu\nu}\varepsilon^{\mu\nu}+
{\bar\xi}^*_\mu{\bar\xi}^\mu+
{\xi}^*_\mu{\xi}^\mu\ .\eqno(4.11)
$$
{}From (4.10$b$) and (4.10$c$) we may eliminate $\bar\xi^\mu$ and $\xi^\mu$ in
(4.11), and express the norm purely in terms of $\varepsilon_{\mu\nu}$,
subject to the constraint implied by (4.10$a$).

     By analogy with the bosonic string, one might expect that the subset of
states of the form (4.9) that are most likely to have negative norm are
those with $\varepsilon_{\mu\nu}$ given by
$$
\varepsilon_{\mu\nu}=\lambda \eta_{\mu\nu}+\kappa \beta_\mu\bar\beta_\nu \ .
\eqno(4.12)
$$
After some algebra, we find that (4.11) can then be written as
$$\eqalign{
{\cal N}&=|\lambda|^2\Big(2\omega_2^{\rm eff}(c^{\rm eff}-3)-c^{\rm eff}+
3(\omega_1^{\rm eff})^2
-6\Big)\cr
&\times \Big(3(1-2\omega_2^{\rm eff})\big((Q^2|\beta_0+\bar\beta_0|^2 -4Q^2
+4Q^2\omega_2^{\rm eff}+(\omega_1^{\rm eff})^2\big)\cr
&+\big(4(\omega_2^{\rm eff}-1)^2
-(\omega_1^{\rm eff})^2\big)(c^{\rm eff}+6Q^2-3)\Big)\ .\cr}\eqno(4.13)
$$
The first factor in parentheses is non-negative for all our effective
intercept values.  From the $L_0^{\rm eff}$ intercept condition (4.5$b$), we
have
$$
Q^2|\beta_0+\bar\beta_0|^2 -4Q^2
+4Q^2\omega_2^{\rm eff}+(\omega_1^{\rm eff})^2\ge0\ .\eqno(4.14)
$$
Thus we find from (4.13) that if $\omega_2^{\rm eff}\le\ft12$, then
$$
{\cal N}\ge|\lambda|^2\Big(2\omega_2^{\rm eff}(c^{\rm eff}-3)-c^{\rm eff}+
3(\omega_1^{\rm eff})^2
-6\Big)\Big((4(\omega_2^{\rm eff}-1)^2
-(\omega_1^{\rm eff})^2)(c^{\rm eff}+6Q^2-3)\Big)\ .\eqno(4.15)
$$
It is easy to see that when $\omega_2^{\rm eff}$ is negative, given by
(3.25), then ${\cal N}\ge 0$.  If $\ell$ in (3.25) takes less than its
maximum value, {\it i.e.}\ $\ell\le n-2$, then ${\cal N}$ is strictly
positive.  One can also check that for $0<\omega_2^{\rm eff}\le \ft12$,
corresponding to a subset of the intercept values given by (3.22), ${\cal
N}$ is again strictly positive.  However for $\omega_2^{\rm eff}>\ft12$,
{\it i.e.}\ the remainder of the intercept values given by (3.22), ${\cal
N}$ can be made arbitrarily negative by taking $|\beta_0+\bar\beta_0|$ to be
large.  Thus there are negative-norm states.

     We have seen above that the complete physical spectrum of the
super-$W_{n+1}$ string includes states with negative norm.  Interestingly
enough, it is amongst the level-1 states with $\omega_2^{\rm eff}>0$ that
the non-unitarity occurs, whilst all the level-1 states with $\omega_2^{\rm
eff}<0$ are unitary.  In order to obtain a unitary theory, it is necessary
to truncate out the negative-norm states in a consistent manner.  Since, as
may be seen from (3.22) and (3.23), the effective spin-1 intercept
$\omega_1^{\rm eff}$ is non-zero whenever $\omega_2^{\rm eff}$ is positive,
it follows that a possible choice for the truncation would be to discard the
physical states associated with all non-zero values of $\omega_1^{\rm eff}$.
Thus we are left with a set of effective $N=2$ superstring theories with
spin-2 intercepts given by
$$
\omega_2^{\rm eff}=-{\ell(\ell+2)\over 4(n+1)},\eqno(4.16)
$$
with $\ell=0,1,\ldots,(n-1)$.

     Since we do not yet have a description of interactions for the physical
states of the super-$W_{n+1}$ string, it is not possible to prove explicitly
that this truncation is consistent.  However, since the effective spin-1
intercept $\omega_1^{\rm eff}$ is a $U(1)$ charge, it seems reasonable to
expect, no matter what the detailed form of the interactions may be, that a
truncation to the states with zero charge should be consistent.  From
(4.5$a$), we see that setting $\omega_1^{\rm eff}=0$ implies that the real
component of momentum $(\beta_0-\bar\beta_0)$ is constrained to be zero.
Since the background charge in the corresponding real direction is zero,
conservation is automatically satisfied for this component of momentum. The
fact that momentum in this direction vanishes for all states means that the
corresponding coordinate is frozen, and hence physically unobservable. Thus
we have the added bonus that although the original theory involved two real
time coordinates, one of them is frozen by the physical-state conditions.
The occurrence of this phenomenon for $N=2$ superstrings was noted in [13].

\bigskip\bigskip
\noindent{\bf 5. Realisations of $N=2$ super-$W_{n+1}$ Algebra}
\bigskip\bigskip

       In section 2, we have given the Miura transformation based on the
$A(n,n-1)$ super algebra, from which one can derive a realisation for $N=2$
super-$W_{n+1}$ in terms of $n$ pairs of complex superfields. The closure of
the algebra requires that the vectors ${\bf \vec H}^{(n)}_j$ in the
differential operator (2.18) satisfy conditions (2.20).  The Miura
transformation given in (2.1) corresponds to a particular choice of the
vectors ${\bf \vec H}^{(n)}_j$ given in (2.21).  This choice of ${\bf \vec
H}^{(n)}_j$ has the nice property that one can express the ${\bf \vec
H}^{(n)}_j$ in terms of ${\bf \vec H}^{(n-1)}_j$, {\it viz.}
$$
\eqalign{
{\bf \vec H}^{(n)}_i&=\Big ( {\bf \vec H}^{(n-1)}_i;\,-1,0 \Big )\ ,\qquad
0 \le i \le 2(n-1)\ ,\cr
{\bf \vec H}^{(n)}_{2n-1}&=\Big (\underbrace{0,0;\,
\ldots;\,0,0}_{2(n-1)};\, -1,1\Big )\ ,\cr
{\bf \vec H}^{(n)}_{2n}&=\Big (\underbrace{0,0;\,
\ldots;\,0,0}_{2(n-1)};\, 0,1\Big )\ ,\cr}\eqno(5.1)
$$
where the ${\bf \vec H}^{(n-1)}_j$ are $2(n-1)$-component vectors also
satisfying (2.20).  It was this property that enabled us to express the
currents of the super-$W_{n+1}$ algebra in terms of those of an arbitrary
super-$W_n$ together with a pair of complex superfields, with explicit
formulae given in (2.9).  Applying this reduction recursively leads to a
realisation of super-$W_{n+1}$ in terms of an arbitrary $N=2$ super
energy-momentum tensor together with $(n-1)$ pairs of complex superfields.

         The above reduction can be easily generalised; one can express the
currents of super-$W_{n+1}$ in terms of those of commuting super-$W_{n-k+1}$
and super-$W_{k}$ algebras together with an additional pair of complex
superfields. To see this, we first express the vectors ${\bf \vec
H}^{(n)}_j$ in terms of ${\bf \vec H}^{(n-k)}_j$ and ${\bf \vec
H}^{(k-1)}_j$, {\it viz.}
$$
\eqalign{
{\bf\vec H}^{(n)}_i&=\Big ({\bf\vec H}^{(n-k)}_i;\,
\underbrace{0,0;\,\ldots;\,0,0}_{2(k-1)};\,-1,0 \Big)\ ,\qquad
0 \le i \le 2(n-k)\ ,\cr
{\bf\vec H}^{(n)}_{2(n-k)+1}&=\Big(
\underbrace{0,0;\,\ldots;\,0,0}_{2(n-k)}; \,
\underbrace{0,0;\,\ldots;\,0,0}_{2(k-1)};\, -1,1\Big)\ ,\cr
{\bf\vec H}^{(n)}_j&=\Big(
\underbrace{0,0;\,\ldots;\,0,0}_{2(n-k)}; \, {\bf \vec
H}^{(k-1)}_{j-2(n-k+1)};\, 0,1\Big )\ , \qquad
2(n-k+1) \le j \le 2n\ ,\cr}\eqno(5.2)
$$
where ${\bf \vec H}^{(n-k)}_j$ and ${\bf \vec H}^{(k-1)}_j$ are the
corresponding vectors for super-$W_{n-k+1}$ and super-$W_k$ respectively,
both satisfying conditions (2.20).  Substituting (5.2) into the differential
operator given in (2.18), we have
$$
\eqalign{
{\cal M}_n&=\prod^{2n}_{j=2(n-k+1)} \Big( \a D+{\bf \vec
H}^{(k-1)}_{j-2(n-k+1)}\cdot D{\bf \tilde\Phi}^\dagger + D\Phi_n\Big)\Big(
\a D+ D\Phi_n-D\bar\Phi_n\Big)\cr
&\times\prod^{2(n-k)}_{j=0}\Big( \a D + {\bf \vec H}^{(n-k)}_j\cdot D{\bf
\vec\Phi}^\dagger-D\bar\Phi_n\Big)\ ,\cr}\eqno(5.3)
$$
where we have grouped the $n$ pairs of superfields ${\bf\vec\Phi}^{(n)}$
into three parts:
${\bf\vec\Phi}\equiv(\Phi_1,\bar\Phi_1;\,\Phi_2,\bar\Phi_2;\,\ldots;\,
\Phi_{n-k},\bar\Phi_{n-k})$, ${\bf\tilde\Phi}\equiv(\Phi_{n-k+1},
\bar\Phi_{n-k+1};\,\Phi_{n-k+2},\bar\Phi_{n-k+2};\, \ldots;
\Phi_{n-1},\bar\Phi_{n-1})$ and an additional pair of superfields
$(\Phi_{n},\bar\Phi_{n})$.  It is straightforward to write ${\cal M}_n$ in
terms of ${\cal M}_{n-k}$ and ${\cal M}_{k-1}$ as follows
$$
{\cal M}_n=e^{-\Phi_n/\a}{\cal M}_{k-1} e^{\Phi_n/\a} \Big( \a D+D\Phi_n
-D\bar\Phi_n\Big) e^{\bar\Phi_n/\a} {\cal M}_{n-k} e^{-\bar\Phi_n/\a}\ .
\eqno(5.4)
$$
Since ${\bf \vec\Phi}$, $\bf \tilde\Phi$ and $(\Phi_n, \bar\Phi_n)$ commute
with each other, we then have a realisation of super-$W_{n+1}$ in terms of
the currents of arbitrary super-$W_{n-k}$ and super-$W_{k-1}$ algebras and
an extra pair of complex superfields.  Similar reductions was also observed
in [15] for $W\!A_n$, $W\!D_n$ and $W\!B_n$ algebras.  Applying this
reduction recursively, we eventually arrive at a realisation of the
super-$W_{n+1}$ algebra in terms of $p$ commuting $N=2$ super
energy-momentum tensors and $(n-p)$ pairs of additional complex superfields,
with $0\le p\le [\ft{n+1}{2}]$.  These super energy-momentum tensors could
be arbitrary except that they must all have the same central charge.  This
follows directly from the fact that the vectors ${\bf \vec H}^{(1)}_j$ for
super-$W_2\equiv$ super-Virasoro ($N=2$) are uniquely determined by the
conditions (2.20) and they give the same Weyl vector
$\vec\bfrho=(-\ft12,-\ft12)$.  Hence all these super energy-momentum tensors
must have the same central charge $c^{\rm eff}$ given by
$$
c^{\rm eff}=3+6\a^2\ .\eqno(5.5)
$$
In this paper we have studied in detail the case of $p=1$.  We construct the
corresponding super-$W_{n+1}$ strings and obtain the complete physical
spectrum.  Investigating the norm of low-lying physical states, we find that
although some physical states are non-unitary, it seems that one can
truncate the physical spectrum into a unitary subsector corresponding to
vanishing $U(1)$ charges.  It will also be interesting to look at the more
general cases, namely super-$W_{n+1}$ strings based on the realisations of
$p$ commuting $N=2$ super energy-momentum tensors and $(n-p)$ pairs of
complex superfields.

\np\bigskip\bigskip
\centerline{\bf REFERENCES}
\frenchspacing
\bigskip\bigskip

\item{[1]}H.\ Lu, C.N.\ Pope, X.J.\ Wang and K.W.\ Xu, ``Anomaly freedom and
realisations for super-$W_3$ strings,'' preprint, CTP TAMU-85/91, to appear
in {\sl Nucl.\  Phys.\ }{\bf B}

\item{[2]}V.A.\ Fateev and S.\ Lukyanov,  {\sl Int.\ J.\ Mod.\  Phys.}
\ {\bf A3} (1988) 507; {\sl Sov.\ Scient.\ Rev.}\ {\bf
A15} (1990); {\sl Sov.\ J.\ Nucl.\ Phys.} \ {\bf 49} (1989) 925.

\item{[3]}A.\ Bilal and J.-L.\ Gervais, {\sl Nucl.\ Phys.}\ {\bf B314}
(1989) 646; {\bf B318} (1989) 579.

\item{[4]}L.J.\ Romans, ``The $N=2$ super-$W_3$ algebra,'' preprint
USC-91/HEP06.

\item{[5]}C.N.\ Pope, L.J.\ Romans and K.S.\ Stelle, {\sl Phys.\
Lett.}\ {\bf 268B} (1991) 167;\nl
{\sl Phys.\ Lett.}\ {\bf 269B} (1991) 287.

\item{[6]}S.R.\ Das, A.\ Dhar and S.K.\ Rama, {\sl Mod.\ Phys.\ Lett.}\
{\bf A6} (1991) 3055;\nl
``Physical states and scaling properties of $W$  gravities and $W$ strings,''
TIFR/TH/91-20.

\item{[7]}H.\ Lu, C.N.\ Pope, S.\ Schrans and K.W.\ Xu, ``The Complete
Spectrum of the $W_N$ String,''  preprint CTP TAMU-5/92, KUL-TF-92/1.

\item{[8]}H.\ Lu, C.N.\ Pope, S.\ Schrans and X.J.\ Wang, ``On Sibling and
Exceptional $W$ Strings,''  preprint CTP TAMU-10/92, KUL-TF-92/8, to appear
in {\sl Nucl.\ Phys.\ }{\bf B}.

\item{[9]}H.\ Lu, C.N.\ Pope, S.\ Schrans and X.J.\ Wang, ``New Realisations
of $W$ Algebras and $W$ Strings,''  preprint CTP TAMU-15/92, KUL-TF-92/11,
to appear in {\sl Mod.\ Phys.\ Lett.\ }{\bf A}.

\item{[10]}D.\ Nemeschansky and S.\ Yankielowicz, ``$N=2$ $W$-algebras,
Kazama-Suzuki models and Drinfeld-Sokolov reduction,'' preprint USC-91/005.

\item{[11]}K.\ Ito, {\sl Nucl.\  Phys.\ }{\bf 370B}(1992) 123.

\item{[12]}D.\ Gepner in ``Superstrings '89,'' (World Scientific 1990).

\item{[13]}H.\ Lu, C.N.\ Pope, X.J.\ Wang and K.W.\ Xu, ``$N=2$ superstrings
with $(1,2m)$ spacetime signature,'' preprint CTP TAMU-99/91, to appear in
{\sl Phys.\  Lett.\ }{\bf B}

\item{[14]}H.\ Lu, C.N.\ Pope and K.S.\ Stelle, ``Massless states in $W$
strings,'' work in progress.

\item{[15]}H.\ Lu and  C.N.\ Pope, `` On realisations of $W$ algebras,''
preprint CTP TAMU-22/92.

\end

\bye